\documentclass[showpacs,preprintnumbers,english,twocolumn,amsmath,amssymb]{revtex4}
\usepackage[T1]{fontenc}
\usepackage[latin1]{inputenc}
\usepackage{graphicx}
\usepackage{amsmath,amssymb}
\usepackage{mathrsfs}
\usepackage{verbatim}
\usepackage{epic,eepic}
\usepackage{babel}
\usepackage{float}
\usepackage{color}

\begin{document}

\def\Journal#1#2#3#4{{#1}, {#2}, {#3}, {#4}.}
\def\ADEP{Advances in High Energy Physics}
\def\ANP{Adv. Nucl. Phys.}
\def\ARNPS{Ann. Rev. Nucl. Part. Sci.}
\def\CTP{Commun. Theor. Phys.}
\def\CPL{Chin. Phys. Lett.}
\def\EPJA{The European Physical Journal A}
\def\EPJC{The European Physical Journal C}
\def\IJMPA{International Journal of Modern Physics A}
\def\IJMPE{International Journal of Modern Physics E}
\def\JCHP{J. Chem. Phys.}
\def\JCP{Journal of Computational Physics}
\def\JHEP{JHEP}
\def\JPCS{Journal of Physics: Conference Series}
\def\JPG{Journal of Physics G: Nuclear and Particle Physics}
\def\NATURE{Nature}
\def\NC{La Rivista del Nuovo Cimento}
\def\NCA{IL Nuovo Cimento A}
\def\NPA{Nucl. Phys. A}
\def\NST{Nuclear Science and Techniques}
\def\PA{Physica A}
\def\PAN{Physics of Atomic Nuclei}
\def\PHY{Physics}
\def\PRA{Phys. Rev. A}
\def\PRC{Physical Review C}
\def\PRD{Physical Review D}
\def\PLA{Phys. Lett. A}
\def\PLB{Physics Letters B}
\def\PLD{Phys. Lett. D}
\def\PRL{Physical Review Letters}
\def\PL{Phys. Lett.}
\def\PREV{Phys. Rev.}
\def\PREP{\em Physics Reports}
\def\PROG{Progress in Particle and Nuclear Physics}
\def\RPP{Rep. Prog. Phys.}
\def\RDNC{Rivista del Nuovo Cimento}
\def\RMP{Rev. Mod. Phys}
\def\SCIENCE{Science}
\def\ZPA{Z. Phys. A.}

\def\ANN{Ann. Rev. Nucl. Part. Sci.}
\def\ANNAST{Ann. Rev. Astron. Astrophys.}
\def\AP{Ann. Phys}
\def\APJ{Astrophysical Journal}
\def\APJS{Astrophys. J. Suppl. Ser.}
\def\EJP{Eur. J. Phys.}
\def\LANC{Lettere Al Nuovo Cimento}
\def\NCA{Nuovo Cimento A}
\def\PHYS{Physica}
\def\NP{Nucl. Phys}
\def\MATH{J. Math. Phys.}
\def\JPAM{J. Phys. A: Math. Gen.}
\def\PRO{Prog. Theor. Phys.}
\def\NPB{Nucl. Phys. B}


\title{Can Tsallis distribution fit all the particle spectra produced at RHIC and LHC?}
\author{ H. Zheng$^{a)}$ and Lilin Zhu$^{b)}$}
\affiliation{
a)INFN, Laboratori Nazionali del Sud, via Santa Sofia, 62, 95123 Catania, Italy\\
b)Department of Physics, Sichuan University, Chengdu 610064, P. R. China}

\begin{abstract}

The Tsallis distribution has been tested to fit the all particle spectra at mid-rapidity from central events produced in d+Au, Cu+Cu, Au+Au collisions at RHIC and p+Pb, Pb+Pb collisions at LHC. Even though there are strong medium effects in Cu+Cu and Au+Au collisions, the results show that the Tsallis distribution can be used to fit most of particle spectra in the collisions studied except in Au+Au collisions where some deviations are seen for proton and $\Lambda$ at low $p_T$. In addition, as the Tsallis distribution can only fit part of the particle spectra produced in Pb+Pb collisions where $p_T$ is up to 20 GeV/c, a new formula with one more fitting degree of freedom is proposed in order to reproduce the entire $p_T$ region. 

\end{abstract}
\maketitle

\section{Introduction}
The heavy-ion collision experiments at RHIC and LHC give us the opportunity to study the phase transition from nuclear matter to quark gluon plasma (QGP), the collective motion, the nuclear medium effects and so on.  The particle spectrum is one of the basic quantities measured in experiments to address the questions raised in such studies. Recently, the Tsallis distribution has attracted the attention of many theorists and experimentalists in high energy heavy-ion collisions \cite{wczhangpp, huapp, khandai2014, li2015ad, phenix201405, wilk201405, liuAuAu2014, pPbAzmi2013, star2007, daupip2006, dauphi2011, phenix2011, alice2,  aliceS2012, cms3, cmsdata7000, sena, cms2014, cleymans, azmiJPG2014,  liAuAu2013, maciej, wongprd, wong2012, wongarxiv2014}. It has been applied to particle spectra produced in different reaction systems, from pp, pA to AA, to understand the particle production mechanism and extract physical quantities, e.g. temperature \cite{huapp, khandai2014, li2015ad, liuAuAu2014, phenix2011, alice2,  aliceS2012, sena, cms2014, cleymans, azmiJPG2014, liAuAu2013, wong2012}, chemical potential \cite{zhaoh20147}. In pp collisions, the excellent ability to fit the spectra of identified hadrons and charged particles in a large range of $p_T$ up to 200 GeV/c is quite impressive \cite{huapp, wongprd, wong2012, wongarxiv2014}. A systematic investigation of particle spectra in p+p collisions at RHIC and LHC has been conducted in ref. \cite{huapp}. The results show that the Tsallis distribution can fit all the particle spectra at different energies in p+p collisions. A possible cascade particle production mechanism is proposed. Recently, a Tsallis distribution scaling function was found for charged hadron spectra in p+p and p+$\bar {\rm p}$ collisions \cite{wczhangpp}.  Comparing to nucleus-nucleus collisions, the pp collision is very simple. It has been used as a baseline for nucleus-nucleus collisions. A nuclear modification factor $R_{pA}$ or $R_{AA}$ was proposed to show the nuclear medium effects in pA or AA collisions referring to pp collisions \cite{phenix201405, daupip2006, dauphi2011, dauk2013, pbpbRAA2012, star2010, auau62pip2007, cucu200pip, auau200pi0, pbpb2760pikp, phenix2011, aliceppb2014}. The nuclear modification factor different from unity is a manifestation of medium effects. Many authors have successfully applied Tsallis distribution to fit particle spectra in pA and AA collisions even though the spectra were affected by nuclear medium modification \cite{khandai2014, li2015ad, phenix201405, wilk201405, liuAuAu2014, pPbAzmi2013, daupip2006, dauphi2011, cms2014, liAuAu2013}. We also notice that many works only show the small $p_T$ part of the particle spectra, while the exponential distribution also can fit the low $p_T$ region \cite{dauphi2011, expstar2009}. It should cover all the $p_T$ regions of particle spectra, available in experiment, in order to show the advantage and/or the fitting power of the Tsallis distribution. In recent years, the experimental groups at RHIC and LHC have published the wide $p_T$ range of particle spectra for different particles in different reaction systems . Such data allow us to conduct the systematic study of particle spectra in heavy-ion collisions at RHIC and LHC using the Tsallis distribution, as we have done for p+p collisions \cite{huapp}. 

In this work, we would like to test whether the Tsallis distribution can fit all the particle spectra produced at RHIC and LHC, which can help us to understand the particle production mechanism. Before we start to conduct our investigation, we can get some clues to estimate whether it can fit the particle spectrum or not from the nuclear modification factor.  If the $R_{pA}$ or $R_{AA}$ is flat for the whole $p_T$ region, according to its definition, this means that the particle spectrum is similar in shape and only differs in magnitude to the one in p+p collisions. Based on the previous studies \cite{huapp, wczhangpp}, we are sure that the Tsallis distribution can fit the particle spectrum since it can fit all the particle spectra produced in p+p collisions, especially up to extremely high $p_T$ \cite{huapp, wongprd, wong2012, wongarxiv2014}. In the pA reactions, $R_{pA}$ are flat and very close to 1 for most of the produced particles at different centralities \cite{phenix201405, daupip2006, dauphi2011, star2010, dauk2013, aliceppb2014}. While in the AA collisions, the nuclear medium effects play an important role. $R_{AA}$ increases from the most central collisions to peripheral collisions \cite{phenix2011, dauphi2011, auau62pip2007, cucu200pip, auau200pi0, pbpbRAA2012, pbpb2760pikp}.  Since the nuclear modification factors of different particles are almost 1 in peripheral heavy-ion collisions \cite{phenix201405, daupip2006, dauphi2011, dauk2013, pbpbRAA2012, star2010, auau62pip2007, cucu200pip, auau200pi0, pbpb2760pikp, phenix2011, aliceppb2014}, as we discussed, Tsallis distribution should be able to fit the particle spectra. Therefore we will only focus on the particle spectra at the most central collisions where the Tsallis distribution may not fit all of them. We have collected data of particle spectra from d+Au, Cu+Cu, Au+Au collisions at RHIC and p+Pb, Pb+Pb collisions at LHC and select the data for most of the particles with the highest $p_T>5$ GeV to conduct this study. 

The paper is organized as following. In section II, we show different versions of the Tsallis distribution used in the literature. More details can be found in ref. \cite{huapp}. We also give the form of Tsallis distribution used in our analysis. In section III, we show our results of particle spectra from d+Au, p+Pb, Cu+Cu, Au+Au and Pb+Pb. Another distribution is proposed to fit the particle spectra in Pb+Pb collisions since Tsallis distribution can only fit part of the particle spectra in the case. A brief conclusion is given in the section IV.  

  \begin{figure} [h]  
        \centering
        \includegraphics[scale=0.4]{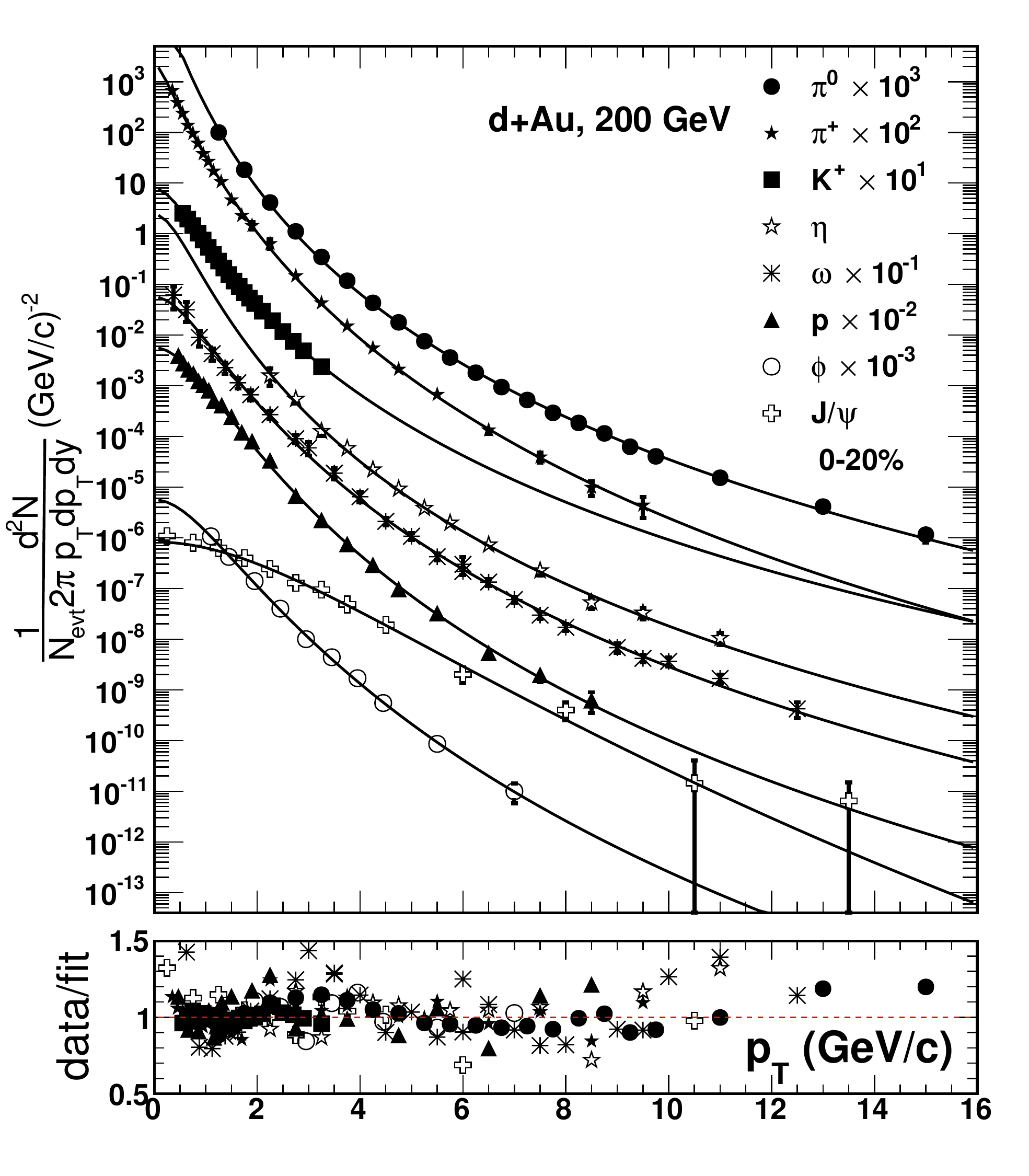}  
\caption{(Color online)The data are from refs. \cite{dau200pi0etaphenix, dau200omegaphenix, dau200jpsiphenix, daupip2006, dauphi2011, dauk2013, dau200gammaphenixadd} for d+Au at $\sqrt{s_{NN}}=200$ GeV. The curves are the analytical results with Tsallis distribution Eq. (\ref{tsallisus}). The corresponding fitting parameters and $\chi^2$/ndf are given in Table. \ref{table1}. For a better visualization both the data and the analytical curves have been scaled by a constant as indicated. The ratios of data/fit are shown at the bottom.} 
\label{figdau}
    \end{figure} 

   \begin{figure} [h]  
        \centering
        \includegraphics[scale=0.4]{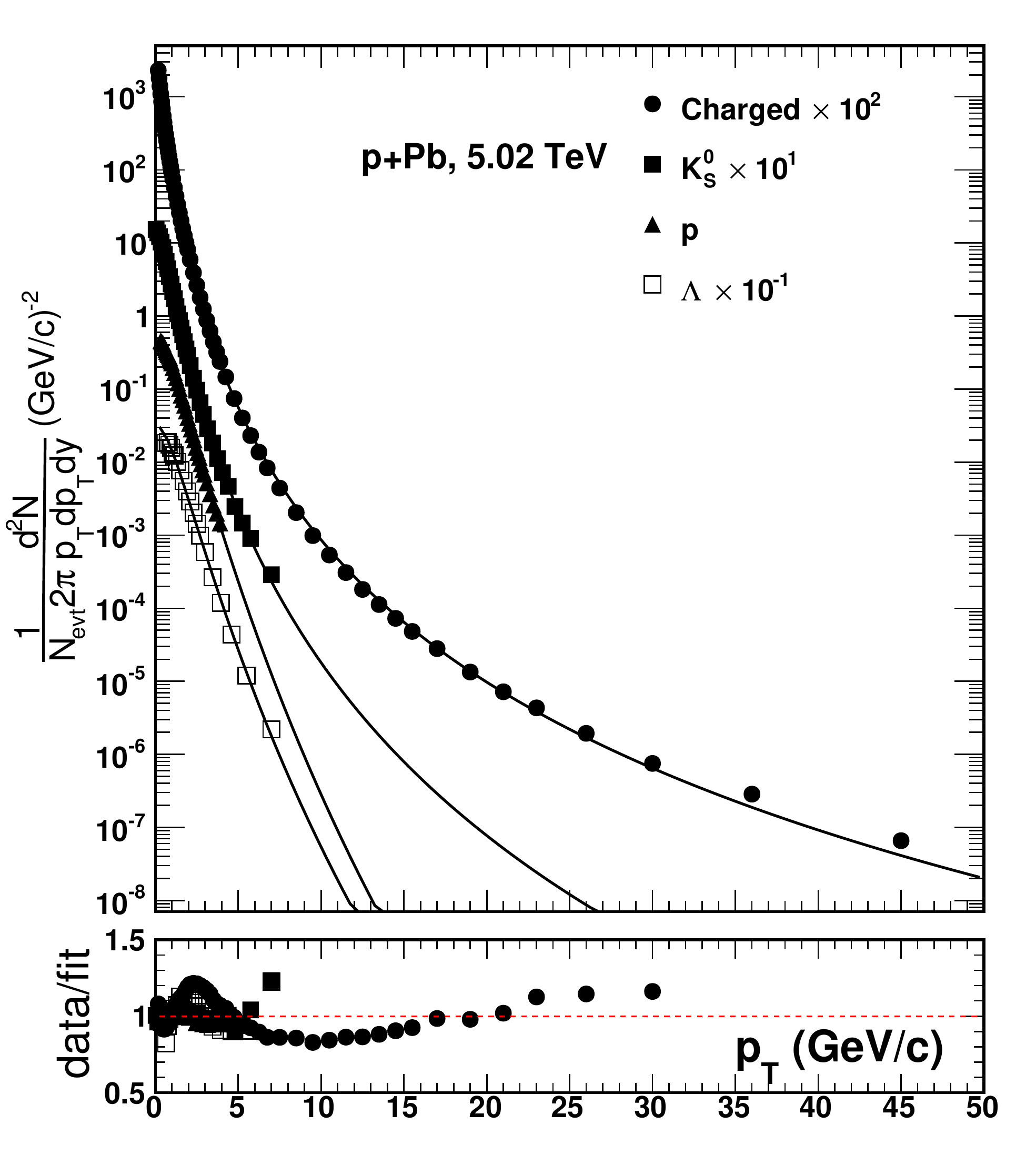}  
\caption{(Color online)The data are from refs. \cite{ppbkplam, aliceppb2014} for p+Pb at $\sqrt{s_{NN}}=5.02$ TeV. The curves are the analytical results with Tsallis distribution Eq. (\ref{tsallisus}). The corresponding fitting parameters and $\chi^2$/ndf are given in Table. \ref{table1}. For a better visualization both the data and the analytical curves have been scaled by a constant as indicated. The ratios of data/fit are shown at the bottom.} 
\label{figppb}
    \end{figure} 

\section{Tsallis distributions}

In the literature, several versions of Tsallis distribution with different arguments can be found \cite{huapp, khandai2014, li2015ad, phenix201405, wilk201405, liuAuAu2014, pPbAzmi2013, star2007, daupip2006, dauphi2011, phenix2011, alice2,  aliceS2012, cms3, cmsdata7000, sena, cms2014, cleymans, azmiJPG2014,  liAuAu2013, maciej, wongprd, wong2012, wongarxiv2014}. The asympotic behaviors of these distributions at low and high $p_T$ limits can be found in \cite{huapp}. We only briefly show them here.

The STAR \cite{star2007}, PHENIX \cite{phenix2011, phenix201405} Collaborations at RHIC along with ALICE \cite{alice2, aliceS2012} and CMS \cite{cms3} Collaborations at LHC adopted this form of Tsallis distribution
\begin{eqnarray}
E\frac{d^3N}{dp^3}&=&\frac{1}{2\pi p_T} \frac{d^2N}{dydp_T} \nonumber\\ 
&=& \frac{dN}{dy} \frac{(n-1)(n-2)}{2\pi nC[nC+m(n-2)]}(1+\frac{m_T-m}{nC})^{-n},\nonumber\\
\label{exptsallis}
\end{eqnarray}
where $m_T=\sqrt{p_T^2+m^2}$ is the transverse mass and $m$ is the mass of the particle.  $\frac{dN}{dy}$, $n$ and $C$ are fitting parameters.

In refs. \cite{cleymans, azmiJPG2014, pPbAzmi2013, liAuAu2013, maciej, liufh20147}, the following Tsallis form  is used
\begin{eqnarray}
E\frac{d^3N}{dp^3} = gV\frac{m_T \cosh y}{(2\pi)^3} [1+(q-1)\frac{m_T\cosh y-\mu}{T}]^{\frac{q}{1-q}}, \label{tsallisB}
\end{eqnarray}
based on thermodynamic consistency arguments. Where $g$ is the degeneracy of the particle, $V$ is the volume, $y$ is the rapidity, $\mu$ is the chemical potential, $T$ is the temperature and $q$ is a parameter. In Eq. (\ref{tsallisB}), there are four parameters $V, \mu, T, q$. $\mu$ was assumed to be 0 in refs. \cite{cleymans, azmiJPG2014, liAuAu2013, liufh20147} which is a reasonable assumption because the energy is high enough and the chemical potential is small compared to temperature. In the mid-rapidity $y=0$ region, Eq. (\ref{tsallisB}) is reduced to
\begin{equation}
E\frac{d^3N}{dp^3} = gV\frac{m_T}{(2\pi)^3} [1+(q-1)\frac{m_T}{T}]^{\frac{q}{1-q}}. \label{tsallisBR}
\end{equation}
In ref. \cite{li2015ad, liAuAu2013}, Eq. (\ref{tsallisB}) has been rewritten as 
\begin{eqnarray}
\frac{dN}{m_Tdm_T} = C\int_{-Y}^{Y}  \cosh y dy m_T [1+(q-1)\frac{m_T\cosh y}{T}]^{\frac{q}{1-q}}, 
\end{eqnarray}
to take into account the width of the corresponding rapidity distribution of the particles.

In ref. \cite{sena}, Sena {\it et al.} applied the non-extensive formalism to obtain the probability of particle with mometum $p_T$ as
\begin{equation}
\frac{1}{\sigma}\frac{d\sigma}{dp_T} = c  p_T\int_0^\infty dp_L \Big[1+(q-1)\beta \sqrt{p_L^2+p_T^2 + m^2}\Big]^{-q/(q-1)},\label{senaeq}
\end{equation}
where $c$ is the normalization constant, $q$ is a parameter, $\beta=\frac{1}{T}$ and $m$ is the mass of particle.
With the approximation $p_T$ very large compared to $p_L$ and $m$ \cite{beck}, Eq. (\ref{senaeq}) can be rewritten as
\begin{eqnarray}
\frac{1}{\sigma}\frac{d\sigma}{dp_T} &=& c[2(q-1)]^{-1/2}B(\frac{1}{2}, \frac{q}{q-1}-\frac{1}{2}) \nonumber \\
&&\times u^{3/2}[1+(q-1)u]^{-\frac{q}{q-1}+\frac{1}{2}},\label{tsallislong}
\end{eqnarray}
where $u=\frac{p_T}{T}$ and B(x, y) is the Beta-function. 

In ref. \cite{wong2012}, Wong {\it et al.} proposed a new form of the Tsallis distribution function to take into account the rapidity cut,
\begin{equation}
(E\frac{d^3N}{dp^3})_{|\eta|<a}=\int_{-a}^{a}d\eta \frac{dy}{d\eta}(\frac{d^3N}{dp^3}). \label{tsalliswong}
\end{equation}
Where
\begin{eqnarray}
\frac{dy}{d\eta}(\eta, p_T)=\sqrt{1-\frac{m^2}{m^2_T\cosh^2 y}}, 
\end{eqnarray}
with
$$y=\frac{1}{2}\ln \Big [\frac{\sqrt{p_T^2 \cosh^2 \eta + m^2}+p_T\sinh \eta}{\sqrt{p_T^2 \cosh^2 \eta + m^2}-p_T\sinh \eta}\Big],$$
and
\begin{equation}
\frac{d^3N}{dp^3}=C\frac{dN}{dy}(1+\frac{E_T}{nT})^{-n},\quad 
E_T=m_T-m, \label{tsalliswongdndp}
\end{equation}
where $C\frac{dN}{dy}$ is assumed to be a constant.

In ref. \cite{huapp}, we have obtained
\begin{equation}
(E\frac{d^3N}{dp^3})_{|\eta|<a} = A(1+\frac{E_T}{nT})^{-n}, \label{tsallisus}
\end{equation}
where $A$, $n$ and $T$ are the fitting parameters. This is equivalent to Eq. (\ref{exptsallis}) but in a simpler form. We adopt Eq. (\ref{tsallisus}) to do the analysis here. We notice that Eq. (\ref{tsallisus}) has been used by the CMS Collaboration \cite{cmsdata7000, cmsB2012} and by Wong {\it et al.} in their recent paper \cite{wongarxiv2014}. The STAR Collaboration has also applied a formula which is very close to Eq. (\ref{tsallisus}) \cite{star2010}.

      \begin{figure} [h]  
        \centering
        \includegraphics[scale=0.4]{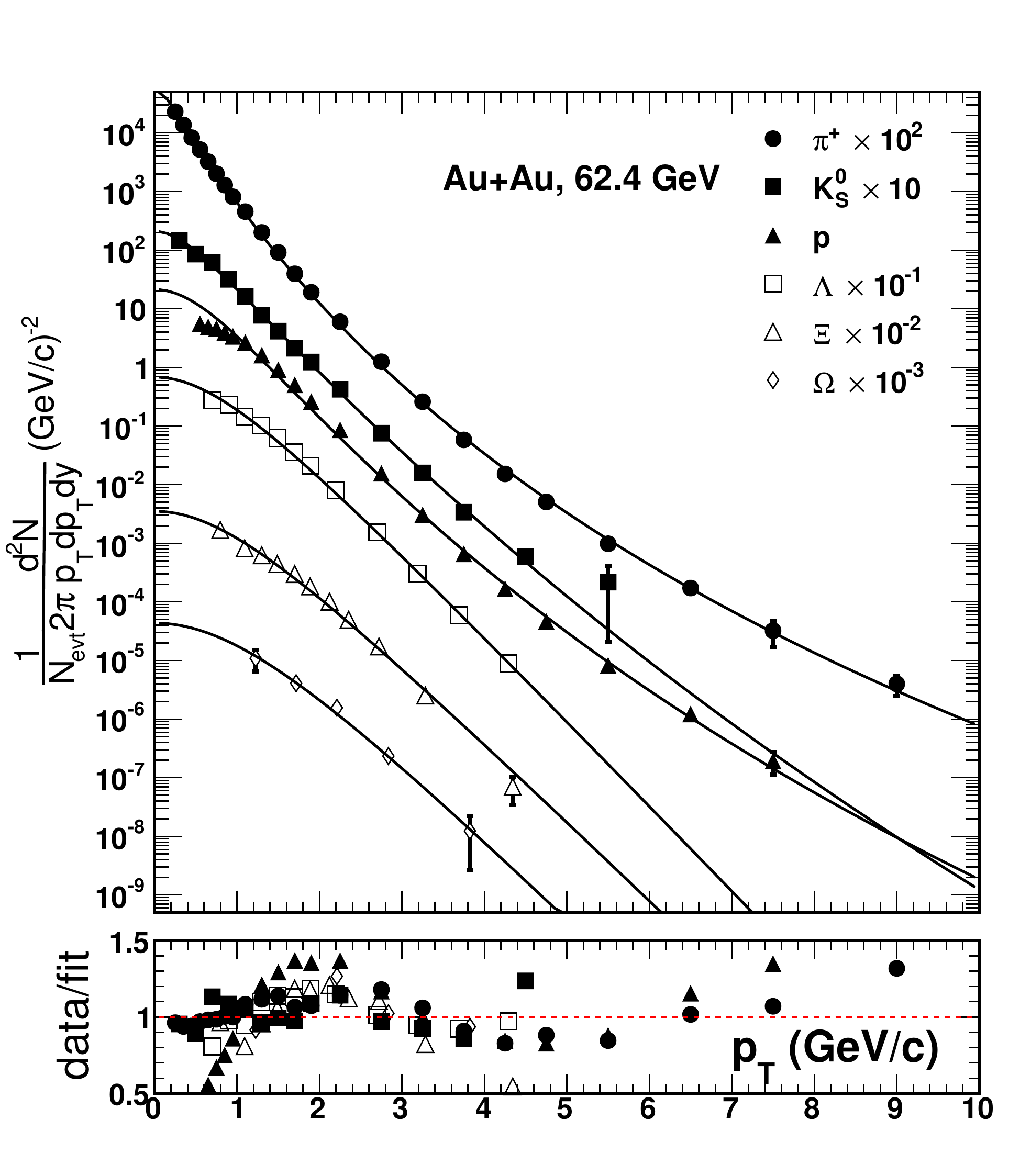}  
\caption{(Color online)The data are from refs. \cite{auau62pip2007, auau62strange} for Au+Au at $\sqrt{s_{NN}}=62.4$ GeV. The curves are the analytical results with Tsallis distribution Eq. (\ref{tsallisus}). The corresponding fitting parameters and $\chi^2$/ndf are given in Table. \ref{table1}. For a better visualization both the data and the analytical curves have been scaled by a constant as indicated. The ratios of data/fit are shown at the bottom.} \label{figauau62}
    \end{figure} 

         \begin{figure} [h]  
        \centering
        \includegraphics[scale=0.4]{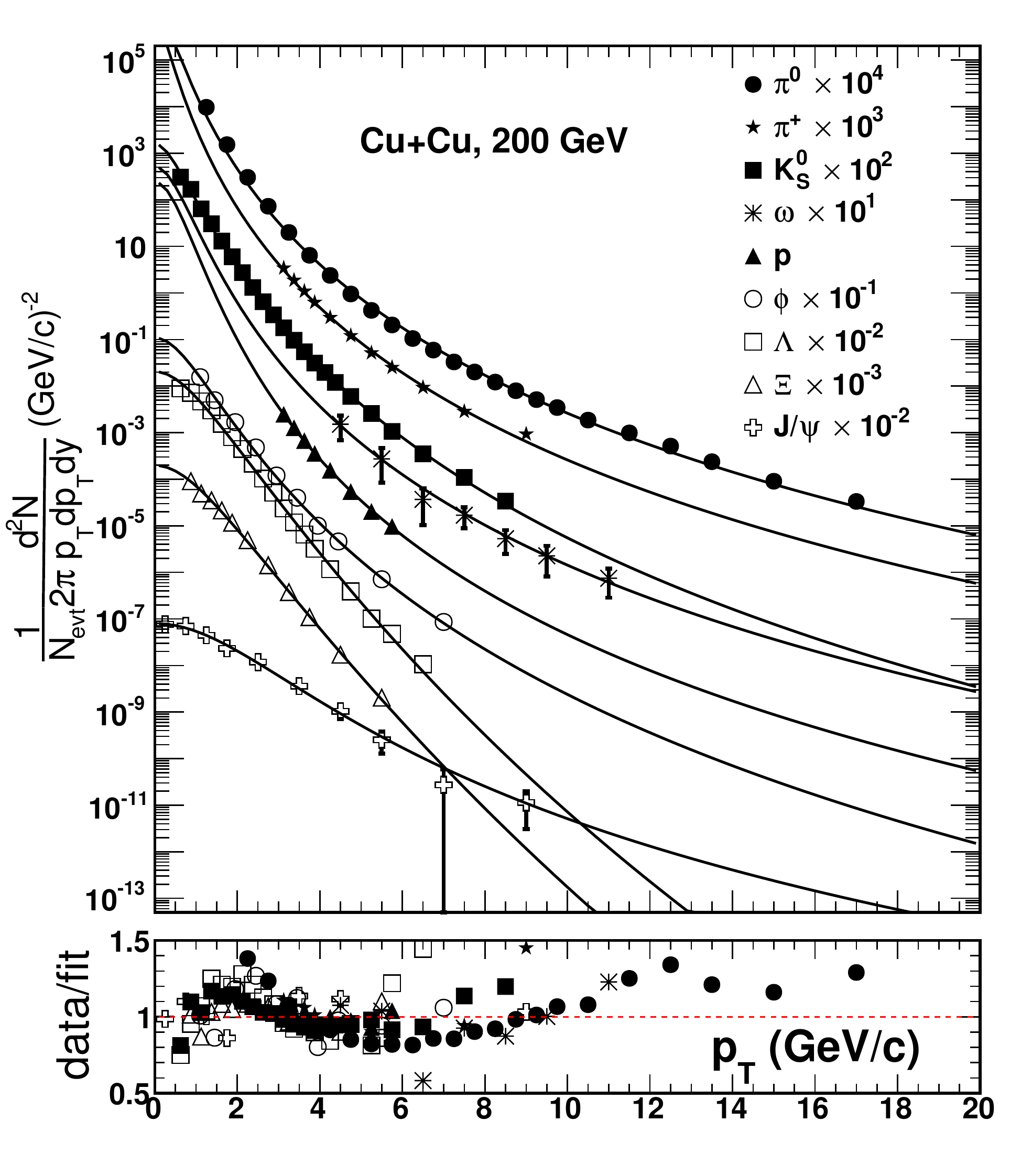}  
\caption{(Color online)The data are from refs. \cite{cucu200pi0phenixadd, dauphi2011, cucu200pip, cucu200klxi, dau200omegaphenix, cucu200jpsiphenixadd} for Cu+Cu at $\sqrt{s_{NN}}=200$ GeV. The curves are the analytical results with Tsallis distribution Eq. (\ref{tsallisus}). The corresponding fitting parameters and $\chi^2$/ndf are given in Table. \ref{table1}. For a better visualization both the data and the analytical curves have been scaled by a constant as indicated. The ratios of data/fit are shown at the bottom.} 
\label{figcucu}
    \end{figure}

\section{Results}
We have selected the data of particle spectra from the most central collisions with the highest $p_T>5$ GeV for most of the particles in d+Au, p+Pb, Cu+Cu, Au+Au and Pb+Pb at RHIC and LHC. We fit the center values of the experimental points. The fit metric used is defined by 
\begin{equation}
M^2 = \sum_{i}[1-\frac{y_i({\rm fit})}{y_i({\rm data})}]^2.
\end{equation}

As we discussed before, the Tsallis distribution should be able to fit the particle spectra from d+Au and p+Pb. One good example has been shown in refs. \cite{li2015ad, phenix201405} for $K_S^0$ and $K^{0*}$ in d+Au at $\sqrt{s_{NN}}=200$ GeV where their spectra can be obtained by multiplying the particle spectra in p+p collisions with $N_{coll}$. In figs. \ref{figdau} and \ref{figppb}, our results for d+Au at $\sqrt{s_{NN}} = 200$ GeV and p+Pb at $\sqrt{s_{NN}} = 5.02$ TeV using Eq. (\ref{tsallisus}) have been shown. In order to see the agreement between  the data and the Tsallis distribution in linear scale, a ratio data/fit is defined. As shown in figs. 1 and 2, the fits for all particles are good. For the left collision systems, we also do the same comparisons. We would like to emphasize that the $p_T$ of charged particle spectrum is up to 45 GeV/c. 

   \begin{figure} [h]  
        \centering
        \includegraphics[scale=0.4]{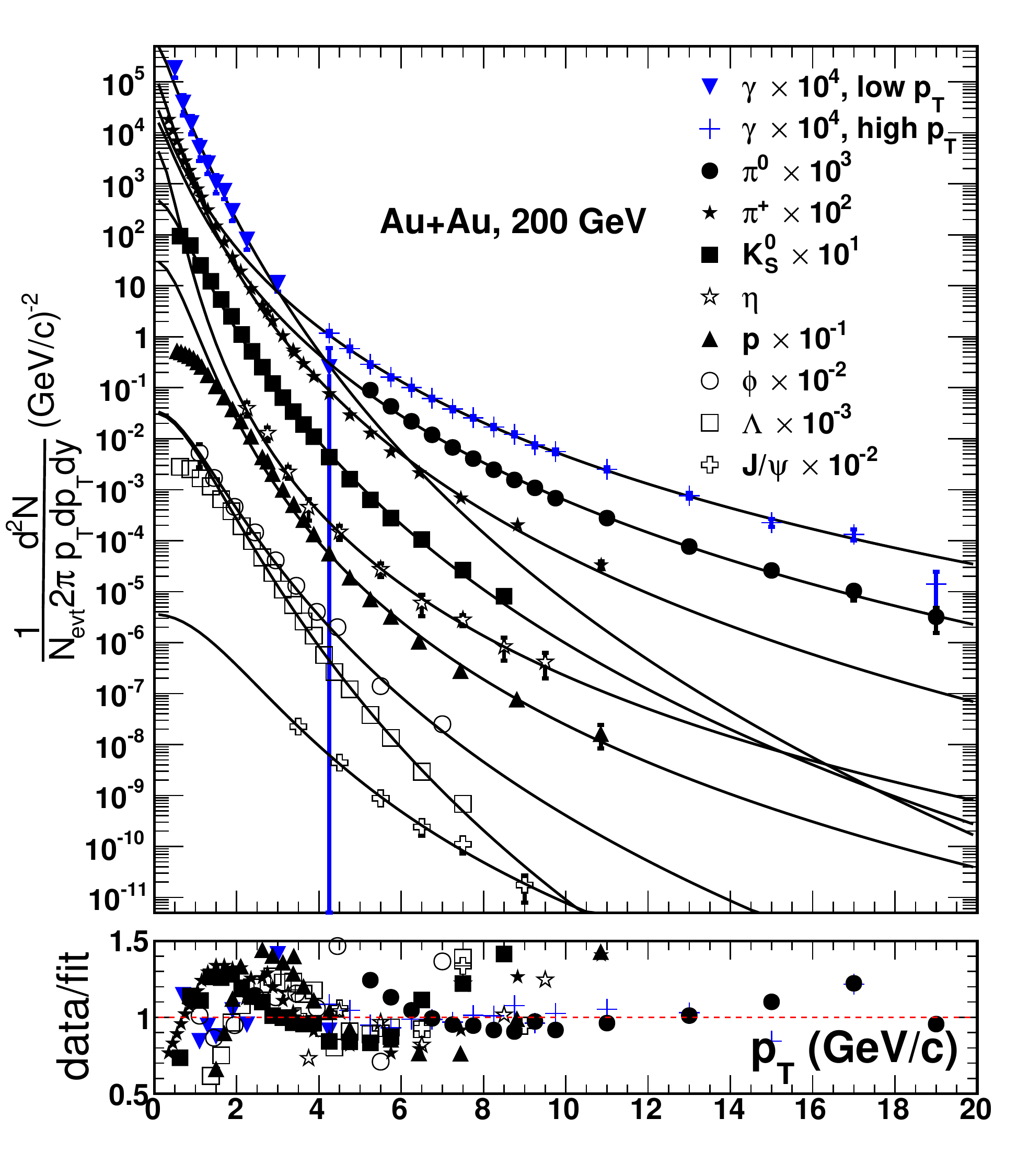}  
\caption{(Color online)The data are from refs. \cite{dauphi2011, cucu200klxi, auau200pi0, auau200pip, auau200etaphenixadd, auau200jpsistaradd, auau200gammahptphenixadd, auau200gammalptphenixadd} for Au+Au at $\sqrt{s_{NN}}=200$ GeV. The curves are the analytical results with Tsallis distribution Eq. (\ref{tsallisus}). The corresponding fitting parameters and $\chi^2$/ndf are given in Table. \ref{table1}. For a better visualization both the data and the analytical curves have been scaled by a constant as indicated. The ratios of data/fit are shown at the bottom.} 
\label{figauau200}
    \end{figure} 
   
     \begin{figure} [h]  
        \centering
        \includegraphics[scale=0.4]{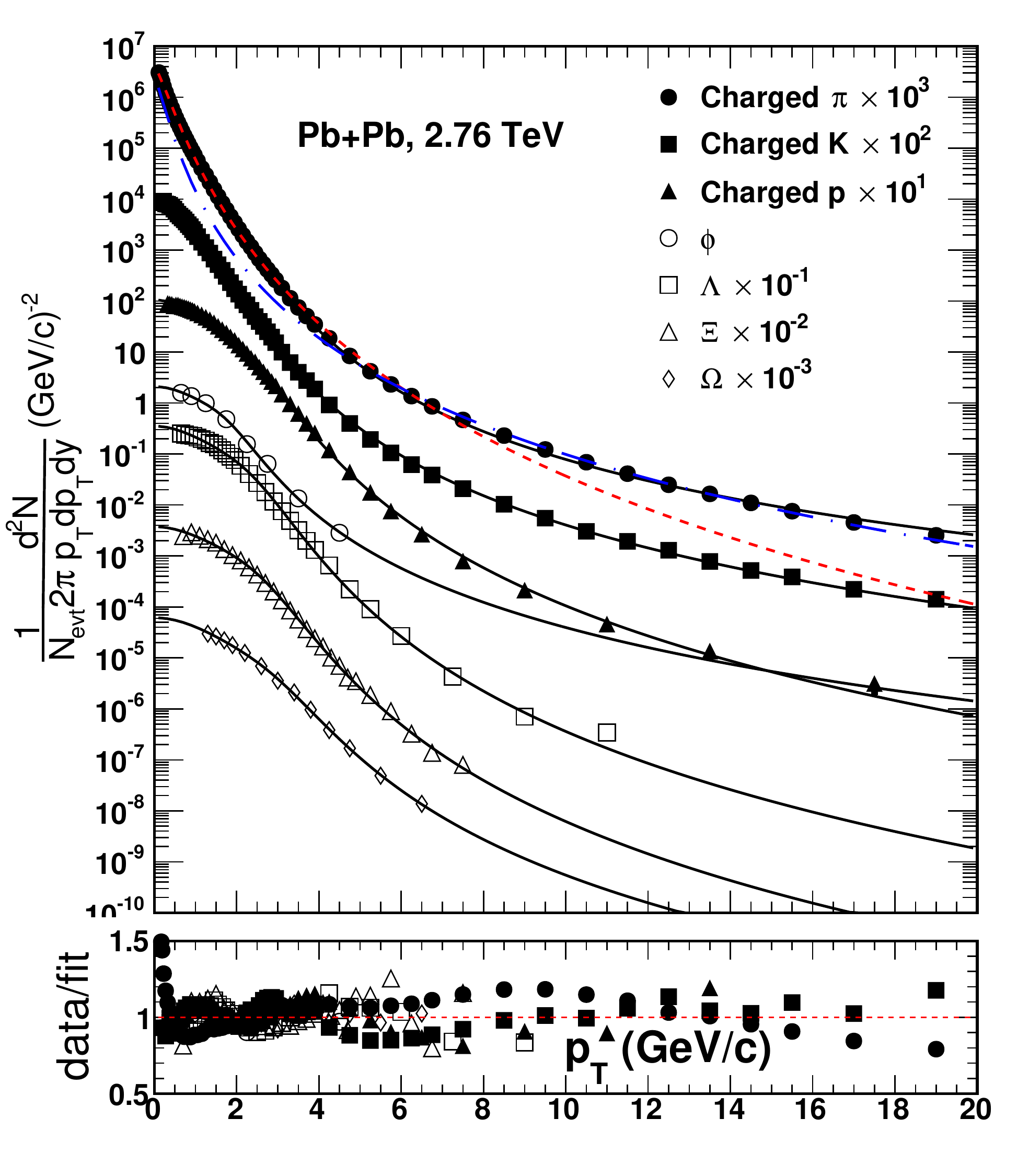}  
\caption{(Color online)The data are from refs. \cite{pbpb2760pikp, pbpb2760phi, pbpb2760lam, pbpb2760xo} for Pb+Pb at $\sqrt{s_{NN}}=2.76$ TeV. The red dashed line is the results fitting the low $p_T$ region and the blue dotted-dashed line is the results fitting the high $p_T$ region using Eq. (\ref{tsallisus}). The solid curves are the analytical results with Eq. (\ref{neweq}). The corresponding fitting parameters and $\chi^2$/ndf are given in Table. \ref{table2}. For a better visualization both the data and the analytical curves have been scaled by a constant as indicated. The ratios of data/fit are shown at the bottom.} 
\label{figpbpb}
    \end{figure} 
    
            \begin{table*}[t]
              \caption{The fitting parameters and the corresponding $\chi^2$/ndf for various particles in different collision systems with Tsallis distribution Eq. (\ref{tsallisus}).} \label{table1}
  \centering
  \begin{tabular}{*{7}{c}}
    \hline
    System & Particle & Centrality & A & T (GeV) & n & $\chi^2$/ndf \\
    \hline
     & $\gamma$ & minimum-bias & 1274.39 & 0.108 & 7.48 & 30.66/21\\
     & $\pi^0$ & 0-20\% & 54.08 & 0.130 & 9.70 &  15.46/18\\
     & $\pi^{+}$ & 0-20\% & 21.60 & 0.173 & 11.56 &   10.74/21 \\
    d+Au & $K^{+}$ & 0-20\% & 0.776 & 0.214 & 8.89  & 1.54/18  \\
    $\sqrt{s_{NN}}=200$ GeV & $\eta$ & 0-20\% & 2.403 & 0.163 & 9.57 & 6.75/10  \\
     & $\omega$ & 0-20\% & 0.565 & 0.224 & 10.53  & 18.08/24  \\
     & $p$ & 0-20\% & 0.568 & 0.221 & 11.97 &  11.60/20 \\
     & $\phi$ & 0-20\% & 0.0563 & 0.270 & 13.63  &  3.57/7\\
     & $J/\psi$ & 0-20\% & 8.30E-7 & 0.582 & 26.91 & 15.67/10 \\
    \hline
       & Charged & minimum-bias & 33.32 & 0.182 & 7.00  & 501.03/57  \\
      p+Pb & $K_s^0$ & 0-5\% & 1.537 & 0.302 & 9.16  &  15.12/31 \\
      $\sqrt{s_{NN}}=5.02$ TeV  & $p$ & 0-5\% & 0.536 & 0.449 & 22.49 & 6.53/36 \\
                & $\Lambda$ & 0-5\% & 0.323 & 0.469 & 19.47 & 21.13/17  \\
   \hline
    & $\pi^{+}$ & 0-10\% & 551.14 & 0.171 & 16.80 & 20.80/20 \\ 
    & $K_s^{0}$ & 0-5\% & 20.03 & 0.264 & 40.88  & 12.95/12 \\
   Au+Au & $p$ & 0-10\% & 21.37 & 0.226 & 22.64 & 305.47/16 \\
    $\sqrt{s_{NN}}=62.4$ GeV& $\Lambda$ & 0-5\% & 6.83 & 0.295 & 11039.30 & 7.26/9 \\
    & $\Xi$ &0-5\% &0.352 & 0.315 & 7811.73 & 6.94/8 \\
    & $\Omega$ & 0-20\% & 4.35E-2 & 0.308 & 3064.05 & 2.57/2 \\
    \hline
    & $\pi^{0}$ & 0-10\% & 304.12 & 0.128 & 9.42 & 38.83/21 \\
    & $\pi^{+}$ & 0-10\% & 1016.57 & 0.111 & 9.45 & 7.16/8 \\
    & $K_s^{0}$ & 0-10\% & 15.89 & 0.198 & 12.19 & 51.37/19 \\
    & $\omega$ & 0-20\% & 51.79 & 0.139 & 9.48 & 1.21/4\\
   Cu+Cu & $p$          & 0-10\% & 242.07 & 0.107 & 9.90 &  0.12/5 \\
   $\sqrt{s_{NN}}=200$ GeV & $\phi$      & 0-10\% & 1.09   &  0.226  & 12.08 & 7.02/8\\
    & $\Lambda$ &  0-10\% & 2.04 & 0.297 & 35.35 &  121.47/16\\
    & $\Xi$        & 0-10\%   & 0.199 & 0.326 & 42.47 & 6.25/7 \\
    & $J/\Psi$   & 0-20\% & 7.98E-6 & 0.399 & 8.16 & 3.60/7\\
    \hline
    & $\gamma$ low $p_T$ & 0-20\% & 109.59 & 0.184 & 19.24 & 4.10/8 \\
    & $\gamma$ high $p_T$ & 0-5\% & 4.64 & 0.187 & 7.85 & 15.44/14 \\
    & $\pi^{0}$ & 0-10\% & 18.68 & 0.191 & 9.06 & 12.36/12 \\
    & $\pi^{+}$ & 0-10\% & 1165.01 & 0.138 & 10.50 & 251.34/28\\
    & $K_s^{0}$ & 0-5\% & 49.56 & 0.213 & 13.99 & 153.73/18 \\
   Au+Au & $\eta$ & 0-20\% & 4978.81 & 0.066 & 8.16 & 4.03/7\\
    $\sqrt{s_{NN}}=200$ GeV & $p$ &0-12\% & 312.81 & 0.118 & 9.50 & 54866.3/26\\
    & $\phi$ & 0-10\% & 3.44 & 0.236 & 13.70 & 7.85/7 \\
    & $\Lambda$ & 0-5\% & 31.74 & 0.215 & 18.91 & 2739.02/18\\
    & $J/\Psi$ & 0-20\% & 3.576E-4 & 0.221 & 7.88 & 0.83/3\\
    \hline
  \end{tabular}
\end{table*}

  \begin{table*}[t]
   \caption{The fitting parameters and the corresponding $\chi^2$/ndf for different particles in Pb+Pb at $\sqrt{s_{NN}} = 2.76$ TeV with Eq. (\ref{neweq}).} \label{table2}
  \centering
  \begin{tabular}{*{8}{c}}
    \hline
    System & Particle & Centrality & A & T (GeV) & b & c & $\chi^2$/ndf \\
    \hline
    & Charged $\pi$ & 0-5\% & 2049.91 & 0.252 & 2.195 & 0.886 & 174.54/59\\
    & Charged $K$ & 0-5\% & 112.68 & 0.346 & 1.776 & 1.150 & 35.37/54 \\
    & Charged $p$ & 0-5\% & 10.55 & 0.710 & 1.845 & 1.605 & 42.29/45\\
   Pb+Pb & $\phi$  & 0-5\% & 2.11 & 0.749 & 1.281 & 1.080 & 1.84/4 \\
    $\sqrt{s_{NN}}=2.76$ TeV & $\Lambda$ & 0-5\% & 3.525 & 0.761 & 1.907 & 1.679 & 23.15/27 \\
    & $\Xi$ & 0-10\% & 0.376 & 0.774 & 2.003 & 1.665 & 32.95/23\\
    & $\Omega$ & 0-10\% & 0.0615 & 0.658 & 2.098& 1.324 & 2.41/9 \\
    \hline
    \end{tabular}
\end{table*}

 Now let us turn to the AA collisions. First we use Tsallis distribution Eq. (\ref{tsallisus}) to fit the particle spectra in Au+Au collisions at $\sqrt{s_{NN}}=62.4$ GeV in fig. \ref{figauau62}. All the particle spectra are well fitted except the proton spectrum at $p_T<1$ GeV/c. This makes a little difference of AA collisions from p+p collisions. We want to check whether this deviation will become larger at higher colliding energy in AA collisions. We considered the particle spectra from Cu+Cu collisions at $\sqrt{s_{NN}}=200$ GeV. The results are shown in fig. \ref{figcucu}. The fitting with Eq. (\ref{tsallisus}) for different particle spectra are very well. But we do not know whether there is deviation or not for proton at low $p_T$ since the data for $p_T<3$ GeV/c are not available. Fortunately, the data for different particle spectra at low $p_T$ in Au+Au collisions at $\sqrt{s_{NN}}=200$ GeV are given. In fig. \ref{figauau200}, one can see the deviations of particle spectra of proton and $\Lambda$ at low $p_T$ from the Tsallis distribution Eq. (\ref{tsallisus}). While a deviation is observed for proton at $p_T<2$ GeV/c which becomes a little larger than the one in Au+Au at $\sqrt{s_{NN}}=62.4$ GeV, all other particle spectra are well fitted. This makes us curious to fit the particle spectra in Pb+Pb collisions at $\sqrt{s_{NN}}=2.76$ TeV. With the successful running at LHC, the identified hadron particle spectra data in Pb+Pb collisions at $\sqrt{s_{NN}}=2.76$ TeV are available up to 20 GeV/c. The data satisfy two criteria. One is that there are strong nuclear medium effects in Pb+Pb collisions which can be seen from $R_{PbPb}$ and the other is that the transverse momenta of the particles reach high values. This gives us an opportunity to test the fitting power of the Tsallis distribution. When we use Tsallis distribution Eq. (\ref{tsallisus}) to fit pion spectrum, we find that Eq. (\ref{tsallisus}) can only fit part of it. If we choose to fit the low $p_T$ region,  Eq. (\ref{tsallisus}) can fit the particle spectrum up to 10 GeV/c, as shown in fig. \ref{figpbpb} with red dashed line. The blue dotted dashed line shows the fit for high $p_T$ region which starts from 4 GeV/c.

 The exponential form equation was used to fit the particle spectra at RHIC when only the low $p_T$ data are available \cite{dauphi2011, expstar2009}. With the upgrade of detectors, we have a better ability to measure the particle spectra.  When the intermediate $p_T$ data are available, the Tsallis distribution is used to understand the particle spectra and extract physical information. The two-Boltzmann distribution was also used in ref. \cite{liuAuAu2014}. But in both cases, the number of free fitting parameters increases from 2 to 3. One fitting degree of freedom is increased in this transition.  In ref. \cite{wilk201405}, a double Tsallis formula was proposed to fit particle spectra obtained from central events in Pb+Pb collisions. In this case, three fitting degrees of freedom are increased. Here we will follow the same logic of the transition from exponential distribution to Tsallis distribution to propose a new form equation to fit the particle spectra in Pb+Pb collisions at $\sqrt{s_{NN}}=2.76$ TeV by only increasing one fitting degree of freedom. We increase the number of free fitting parameters from 3 to 4 and the proposed formula is
 \begin{equation}
 (E\frac{d^3N}{dp^3})_{|\eta|<a}=A \frac{e^{-\frac{b}{T}\arctan(E_T/b)}}{[1+(\frac{E_T}{b})^4]^{c}}. \label{neweq}
 \end{equation}
There are four parameters A, b, T and c. We are inspired by the solution of Fokker-Planck equation \cite{fokker}. We change the power from 2 in ref. \cite{fokker} to 4 in Eq. (\ref{neweq}) in order to fit well all the particle spectra with one equation. Fig. \ref{figpbpb} shows the fits with Eq. (\ref{neweq}) are excellent. We would like to mention that when $\frac{E_T}{b}<<1$, Eq. (\ref{neweq}) becomes
\begin{equation}
(E\frac{d^3N}{dp^3})_{|\eta|<a}\propto e^{-\frac{E_T}{T}},
\end{equation} 
and when $\frac{E_T}{b}>>1$, 
\begin{equation}
(E\frac{d^3N}{dp^3})_{|\eta|<a}\propto p_T^{-4c}.
\end{equation} 
Eq. (\ref{neweq}) has the same asymptotic behaviors as Eq. (\ref{tsallisus}).

\section{Conclusions}
In this paper, we have tested the fitting ability of Tsallis function by fitting different particle spectra produced at the most central collisions in d+Au, p+Pb, Cu+Cu, Au+Au and Pb+Pb at RHIC and LHC. The Tsallis distribution is able to fit all the particle spectra in d+Au and p+Pb collisions where the medium effects are very weak. This information can be obtained by the nuclear modification factor. In the AA collisions, the Tsallis distribution can fit all the particle spectra very well at RHIC energies except the little deviation observed for proton and $\Lambda$ at low $p_T$.  However the Tsallis distribution can only fit part of the particle spectra in Pb+Pb at $\sqrt{s_{NN}}=2.76$ TeV, either in the low or high $p_T$ region. We have proposed a new formula in order to fit all the particle spectra in Pb+Pb by increasing one fitting degree of freedom from Tsallis distribution. This follows the same idea of the transition from the exponential distribution to Tsallis distribution when intermediate $p_T$ data are available in experiments. 

According to the results in this paper and ref. \cite{huapp}, we conclude that we can do the systematic analysis of particle spectra with Tsallis distribution in p+p, pA at RHIC and LHC. In the AA collisions at $p_T<10$ GeV/c, we can do the same analysis as in p+p and pA at RHIC and LHC. But when we consider Pb+Pb collisions, the Tsallis distribution fails.

\section*{Conflict of Interests}
The authors declare that there is no conflict of interests regarding the publication of this paper.

\section*{Acknowledgments}
We thanks Dr. J. Mabiala for reading carefully our manuscript. This work was supported,  in part,  by the NSFC of China under Grant No.\ 11205106.

\end{document}